\documentclass[fleqn,twoside]{article}
\usepackage{espcrc2}

\usepackage{graphicx}
\usepackage[figuresright]{rotating}

\newcommand{\AmS}{{\protect\the\textfont2
  A\kern-.1667em\lower.5ex\hbox{M}\kern-.125emS}}

\title{Small x divergences in a heavy quark-antiquark state.}

\author{Martina Brisudov{\'a}\address{Institute of Physics, Slovak Acad. Sciences,  
        D{\'u}bravsk{\'a} cesta 9, Bratislava 842 28, Slovakia }%
        \thanks{Present address: NTC, Indiana Univ., 2401 Milo B. Sampson Lane,
         Bloomington, IN 47405, USA.
         {\it E-mail address:} brisuda@niobe.iucf.indiana.edu} }

\begin{document}

\begin{abstract}
With the current state of similarity renormalisation group approach to light-front QCD,
 it is possible to address with a degree of generality the issue of light-cone zero modes. 
 We find, contrary to earlier results in a less general framework,
  that infrared divergences associated with the zero
  modes do not cancel out in a color singlet heavy quark-antiquark states, except for 
 the lowest order in the nonrelativistic expansion.
\vspace{1pc}
\end{abstract}

\maketitle

\section{Introduction}

\subsection{Light-cone zero modes}

Light-cone zero modes are surely the most controversial subject in light-cone 
field theories. Practically everybody in this field worked on this at some point, and
I apologize that I cannot mention everyone for the lack of space.
Nevertheless, however extensive the literature on light-cone zero modes, there are basically 
two attitudes  to this issue. One is to try to explicitly solve for the zero modes. 
This is typically done for various, often lower
 dimensional, field theories (for 
review and references see \cite{BrodskyPauliPinsky}, for extensive list of references see 
also \cite{thelongpaper}) in the context of discreet 
light cone quantization (DLCQ) (for a recent review of the method as well as excessive 
list of references see \cite{DLCQ}). 

 I would like to mention a somewhat untypical paper that falls into this cathegory, a
recent work by Tomaras, Tsamis and Woodard \cite{woodard} on 
back reaction in light-cone QED. Though motivated by the back-reaction in quantum 
gravity occuring on an inflating background, their work addresses some issues of 
the light-front vacuum without having to evoke DLCQ.  They consider 
 a free QED  coupled to a constant external electric field 
in  continuum (3+1) dimensions, and a full operator solution to the model is constructed. 
In this set up, 
all modes are {\it forced} to go through 
the zero mode at which point particle pairs are created. The zero mode of the constraint
components of the fermionic field is shown to be crucial for unitarity.

The other approach to the problem of zero mode is more pragmatic. Instead of 
trying to solve for the zero mode, it is simply cutoff, be it with DLCQ 
\cite{joelthorn}
 or an explicit infrared cutoff in a continuous formulation 
\cite{thelongpaper}. Physics associated with this mode can then be put in 
form of counterterms, if needed, for example, to restore symmetries or account 
for phenomena associated with the vacuum. Traditionally, spontaneous symmetry 
breaking was viewed as an example of such a phenomenon. 
However, Rozowsky and Thorn \cite{joelthorn} 
have argued recently that, while conceding that the inclusion of a 
fundamental zero mode is a valid 
theoretical option, it is not necessary to describe spontaneous symmetry 
breaking where its presence seems to be most needed.  Indeed, in scalar 
quantum 
field theory in (1+1) dimensions DLCQ the physics of spontaneous symmetry breaking is 
completely and accurately described without the zero modes \cite{joelthorn}.

I hope that the two examples I mentioned explicitly are sufficient to 
remind you how confusing and
 controversial the zero modes are.

\subsection{Similarity renormalisation group approach to light-cone QCD}
The similarity renormalisation group approach has been presented 
by various authors at the light-cone conferences many times since its introduction  
\cite{thelongpaper}.
 The basic assumption upon which the approach is based is
that it is possible to {\it derive a constituent picture for hadrons from
QCD} \cite{thelongpaper}.  If this is possible, nonperturbative bound 
state problems in
QCD can be approximated as coupled, few-body Schr{\"o}dinger equations.

The starting point is a regulated canonical light-front Hamiltonian.  The apparent difficulties
 with renormalization of light-front Hamiltonians (compared to
Lagrangians) are turned into an advantage by using  similarity renormalization 
which allows to
transform the standard perturbative QCD Hamiltonian at high energy scales 
into an effective Hamiltonian with formfactors restricting momentum transfers at hadronic  
 scales. Unfortunately, renormalisation group as we presently know it can systematically
remove dependence only on one regulator. That leaves the infrared, or small $k^+$,
 regulator in the game.

The need to put in new counterterms associated with the infrared (IR) regulator in our 
approach  was 
anticipated \cite{wilsonja} but was not  encountered yet in the applications to hadronic
physics so far \cite{us}. Perry \cite{P2} has shown that even though the one 
body and two-body effective operators are each separately divergent as $k^+$ 
goes to zero, the divergences {\it exactly} cancel in any color singlet 
state. The cancelation does not occur for non-singlet states, leaving them 
with an infinite mass. This together with a naturally generated confining 
potential (imprecisely referred to as ``logarithmic'') is a plausible feature of 
the approach. Note that both the effective confining potential and the 
infinite mass of the color non-singlets originate from small $k^+$ regions.
  
A shortcoming of the above described result is that it was obtained in a 
similarity renormalization group limited to matrix elements that required a 
specific infrared regulator (theta function) and an introduction of an arbitrary scale 
${\cal P}^+$ which violates an explicit kinematic symmetry of light cone. In a 
bound state calculation one can argue that there is a preferred scale, i.e. 
one associated with the typical longitudinal momentum of the state, or, the total center of
 mass $P^+$; however, consequences of 
the violation of the kinematic symmetry  are not known.

Since then, the similarity renormalization group approach has advanced so that 
it is no longer necessary to violate the kinematic boost invariance, and to generate
 counterterms dependent on the total center of mass $P^+$ (see, for example \cite{gluons}). 
 It is also possible to use an arbitrary form of 
the small $x$ regulator  ($x$ being a dimensionless, boost invariant fraction
of the total $P^+$). Thus, for the first time we are able to study with some degree of
generality the issues related to the light-cone zero mode. 
Recently, G{\l}azek has found \cite{gluons}
that even though infrared divergent terms cancel out in the running coupling, there is a
residual finite dependence on the functional form of the infrared regulator.  
We wish to study 
the issue of small $x$ (or infrared) divergences in color singlet states 
consisting of a heavy quark and antiquark of the same flavor, for simplicity.
Does the cancelation found by Perry occur also in the boost invariant formulation?

\section{Cancelation of infrared divergences?}
The short answer to this question is: No. Details of this calculation can be found in
\cite{ja}. Here we just show the resultant bound state equation for the binding $E$ 
and wavefunction $\Phi_{12}~\equiv~\Phi (x_1, x_2=1-x_1, \kappa_{12}^{\perp})$, 
(all other symbols will be explained below),

\begin{eqnarray}
\lefteqn{\left( 4m^2 + 4m \, E\right) \Phi_{12}=}\nonumber\\
 \lefteqn{{\kappa_{12}^2 +m^2(\lambda) \over{x_1 x_2}} \Phi_{12} 
  - {g^2 \over{4 \pi^2}} C_F \times}\nonumber\\
 \lefteqn{\left\{
\int {{\rm d}x_3 {\rm d}^2\kappa_{34}^{\perp} \over{\pi}}
\left[ {\cal V}_1\!+\! {\cal V}_2\! +\! {\cal V}_{\rm inst} \right]f_{\lambda}({\cal 
M}_{12}^2 -{\cal M}_{34}^2) \Phi_{34} \right. }\nonumber\\
& &  \left. - {\sqrt{\pi} \lambda^2 
\over{\sqrt{x_1^2+x_2^2}}} 
\int_0^{1} {{\rm d}y \over{y}}r_{\delta}\left( x_1 \, y\right)
r_{\delta}\left( x_2 \, y\right)  \Phi_{12} \right\} \nonumber\\
\lefteqn{+{g^2\over{4 \pi^2}}\, 
C_F \, {\sqrt{\pi}\over{2}}\, \lambda^2 
{\cal I}_{\delta}(x_1,x_2)\Phi_{12}\   ,}\label{boundsteqn2}
\end{eqnarray}
with

\begin{eqnarray}
\lefteqn{{\cal I}_{\delta}(x_1)\equiv }\nonumber\\
&+&{1\over{\sqrt{2} \ x_1 x_2}} 
 \int_0^1 \, {{\rm d}y\over{y}}\left(r_{\delta}(y)\right)^2  \nonumber\\
 & - &
 {2\over{\sqrt{x_1^2+x_2^2}}}
\int_0^{1} {{\rm d}y \over{y}}r_{\delta}\left( x_1 \, y\right)
r_{\delta}\left( x_2 \, y\right). \label{i}
\end{eqnarray}

Here $m$ is current (heavy) quark mass, $m(\lambda)$ is quark mass including 
finite contribution from effective one-body operators at finite $\lambda$, similarity scale. 
$\kappa_{ij}^{\perp}$ is the standard relative transverse momentum between particles $i$
 and $j$. ${\cal M}_{ij}^2= (p_i +p_j)^2$ is the invariant mass of the state consisting of
 particles $i,\ j$. $f_{\lambda}$ is the similarity formfactor, ${\cal V}_1, \ {\cal V}_2$ are the
coefficients of the effective two-body operator at order $g^2$ and  ${\cal V}_{\rm inst}$ is the
instantaneous interaction (for details see \cite{ja}). Finally, $r_{\delta} (y)$ is a 
(general) infrared regulator satisfying 
${ \lim_{\delta \rightarrow 0} r_{\delta} (y) =1}$, and $\delta$ is the IR cutoff. 

The apparently $\delta$-dependent term in curly brackets in  (\ref{boundsteqn2}) is in fact
 ensuring that the integral part of the bound state equation is independent of $\delta$. 
 All $\delta$-dependence is contained in the last line of the bound state equation, i.e. in 
 the $x_1$ dependent ${\cal I}_{\delta}(x_1)$ given in (\ref{i}). 
If ${\cal I}_{\delta}(x_1)$ vanished identically for all values of 
$x_1$, the bound state equation (\ref{boundsteqn2}) would be
 independent of $\delta$, just as in Hamiltonian matrix element formulation of the 
 similarity renormalization \cite{P2}.
 
It is obvious that this does not happen here, except for the leading order in the
 nonrelativistic expansion, i.e. the point 
$x_{i} = 1/2$. However, an arbitrarily small deviation from $x=1/2$ 
introduces a positive divergent constant into the bound state equation in the 
limit $\delta \rightarrow 0$.
 The bound state equation therefore is not defined in this limit.
 
 Is there a solution to this problem within the framework of the (IR) cutoff theory?
 Do we need to solve for the zero modes first? These questions are open at present.
 Hopefully, by the next light-cone meeting we will have answers. 

\vskip.1in

I would like to thank the organizers for hospitality during the workshop, 
and also for the compassion and care extended to us on September~11.

\end{document}